\documentclass[preprint,aps]{revtex4} 

\usepackage{amsmath}
\usepackage{graphicx}

\def\be{\begin{equation}}
\def\ee{\end{equation}}
\def \bea#1\eea {\begin{eqnarray}#1\end{eqnarray}}

\begin{document}


\title{Structure of the First and Second Neighbor Shells of Water: Quantitative Relation with Translational and Orientational Order}
\author{Zhenyu Yan$^1$, Sergey V. Buldyrev$^{2,1}$, Pradeep Kumar$^1$,
Nicolas~Giovambattista$^{3}$, Pablo G. Debenedetti$^{3}$ and H. Eugene Stanley$^1$}

\affiliation{
$^1$Center for Polymer Studies and Department of Physics, Boston
University, Boston MA 02215, USA\\
$^2$Department of Physics, Yeshiva University, 500 West 185th Street,
New York, NY 10033 USA\\
$^3$Department of Chemical Engineering, Princeton University,
Princeton, New Jersey 08544-5263 USA
}

\date{\today}

\newpage

\begin{abstract}

We perform molecular dynamics simulation of water using the TIP5P model to quantify
structural order in both the first shell (defined by four nearest neighbors)
and second shell (defined by twelve next-nearest neighbors) of a central water molecule.
We find the anomalous decrease of orientational order upon compression occurs in both shells, 
but the anomalous decrease of translational order upon compression occurs {\it mainly in the second shell}.
The decreases of translational and orientational orders upon compression (``structural anomaly")
are thus correlated only in the second shell. Our findings quantitatively confirm
the qualitative idea that the thermodynamic, dynamic and structural anomalies of water
are related to changes in the second shell upon compression.

\end{abstract}


\maketitle

It is known both from experiments~\cite{
bosio83, soper0002, bottu02, strassle06} and simulations~\cite{canpolat98,Schwegle00, ssitta03}
that the first shell of a central water molecule, usually defined by the first minimum of the oxygen-oxygen (O-O)
pair correlation function (PCF), can accomodate between four and five water
molecules, depending on pressure~\cite{scio90}. The signature of this first shell, defined by the first
maximum of the O-O PCF, barely changes with pressure. In contrast, the properties of the
second shell, which extends between the first and second minima of the O-O PCF, are highly
dependent on pressure, indicating that large structural changes occur in this shell
upon compression~\cite{soper0002}.

The structural order of water has been quantified by two 
measures~\cite{jeffrey01}: a local {\it orientational} order parameter $q$, which quantifies
the extent to which a molecule and its four nearest neighbors
adopt a tetrahedral local structure in the first shell,
and a {\it translational} order parameter $t$, which quantifies the tendency of molecular
pairs to adopt preferential separations. While $q$ depends only on the four
nearest neighbors of a central molecule in its first shell,
$t$ depends on all the neighbors of the central molecule.

Water in the \emph{liquid} phase displays: 
(i) a thermodynamic anomaly (density decrease upon cooling or, equivalently, entropy increase with pressure);
(ii) a dynamic anomaly (increase of diffusivity upon compression);
(iii) a structural anomaly (decrease of both $q$ and $t$ upon compression)~\cite{jeffrey01}.
Several other liquids with local tetrahedral 
order~\cite{angellPCCP,sri03} such as silica, silicon, carbon and phosphorous
also show waterlike anomalies. 
In the case of water~\cite{jeffrey01}
and silica~\cite{shell02}, computer simulation studies show that the
anomalies (i)-(iii) in the liquid phase are closely related. For
example, in the case of water, the region of thermodynamic anomaly
in the temperature-density ($T$-$\rho$) plane
is enclosed by the region of dynamic anomaly, which in turn is enclosed by 
the region of structural anomaly~\cite{jeffrey01}.

Recent studies show that simple liquids
interacting via spherically-symmetric potentials 
can exhibit waterlike anomalies~\cite{Jagla99,zyan,predp05},
suggesting that strong orientational interactions in the first shell 
are not \emph{necessary} for a liquid to show thermodynamic, dynamic and structural anomalies.
In light of these findings, it remains unclear how much the 
strongly orientation-dependent first-shell interactions and the second-shell 
environment each contribute to water's anomalies. To address these questions,
we first modify the definition of first and second shells for the purpose of quantitative study.
Then we define the orientational and translational order parameters within each shell and study their
changes with $T$ and $\rho$.

\begin{figure}
\includegraphics[width=10cm]{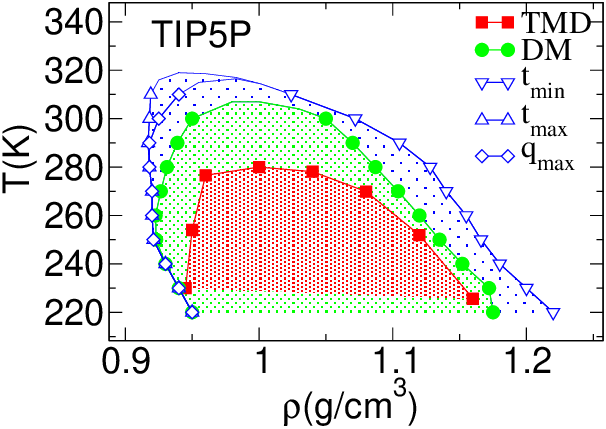}
\caption{
(Color online) Three anomalous regions in $T$-$\rho$ plane for 512 molecules interacting with TIP5P potential.
(i) The density anomaly region is defined by the locus of density maxima (TMD),
inside of which the density increases when the system is heated at constant pressure.
(ii) The diffusion anomaly region is defined by the loci of diffusion maxima or minima (DM),
inside which the diffusivity increases with density.
(iii) The structural anomaly region is defined by the loci of translational order 
minima ($t_{\rm min}$) and maxima ($t_{\rm max}$), or orientational order maxima ($q_{\rm max}$), 
inside which both translational and orientational orders decrease with density
(see Fig.~\ref{tq-d}).
}
\label{tmddmqt}
\end{figure}

We perform constant volume isothermal (NVT) molecular dynamics
simulation of $512$ TIP5P (five-site transferable interaction potential) water molecules.
Our simulations are performed using a cubic box with periodic boundary 
conditions. We control the temperature using a Berendsen thermostat~\cite{berendsen84}.
The TIP5P model reproduces the thermodynamic properties of water
over a broad region of the phase diagram~\cite{tip5p}. 
In particular, we find that the TIP5P model reveals similar relations between
the thermodynamic, dynamic and structural anomalies as observed in ref.~\cite{jeffrey01}
using the SPC/E model (see Fig.~\ref{tmddmqt}).

\begin{figure}
\includegraphics[width=10cm]{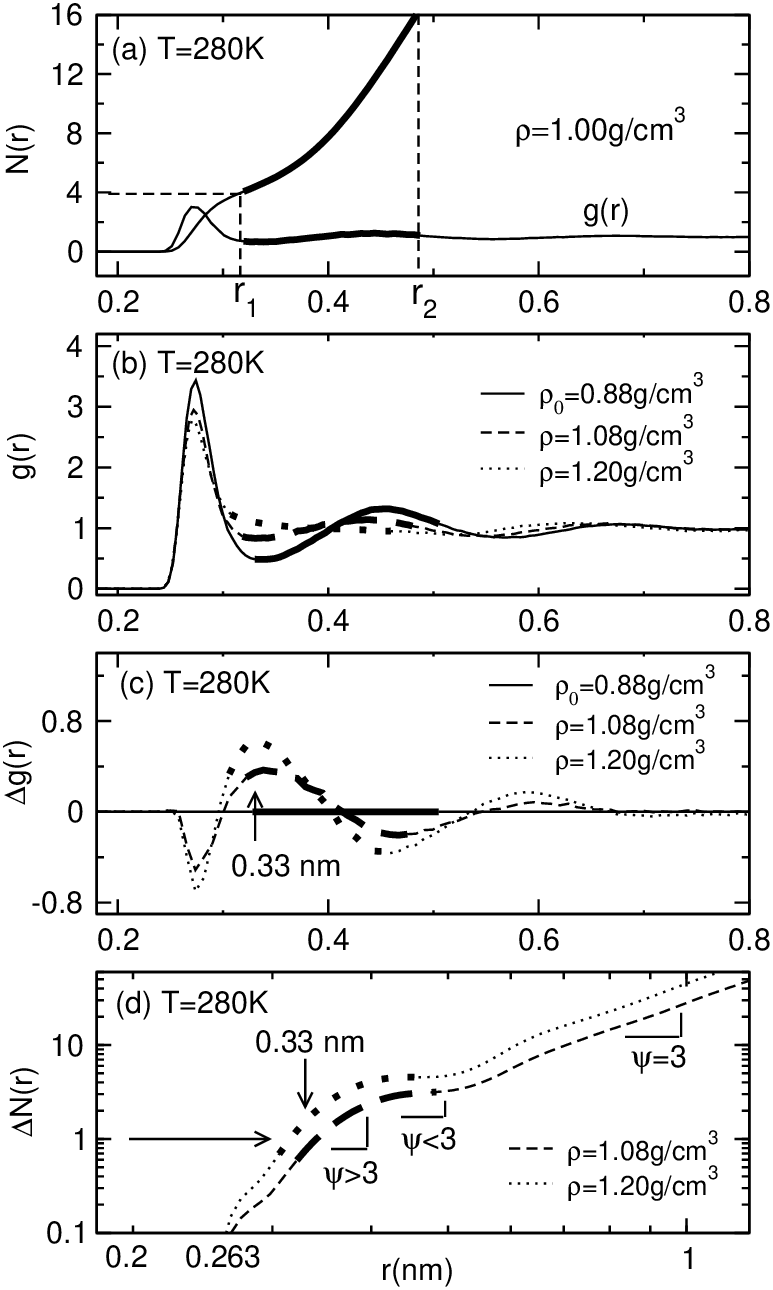}
\caption
{
(a) The number of neighbors $N(r)$ around a central water molecule.
$r_1$ and $r_2$ correspond to the first and the second shell distances, defined such that $N(r_1)=4$ and $N(r_2)=16$.
(b) The O-O PCF $g(r)$, (c) Difference $\Delta g(r)$ between $g(r)$ at a given density and
$g(r)$ at $\rho_0$, and (d) Difference $\Delta N(r)$ between $N(r)$ at a given
density and $N(r)$ at $\rho_0$ for TIP5P water. $\psi$ characterizes
the local slope. The bold portions of the curves correspond to water's
second shell, $r_1 < r \leq r_2$, showing that the largest changes upon compression occur in the second shell.
}
\label{rdf}
\end{figure}

The first and second shells of water can be defined according to the first and second minima of
the PCF, $g(r)$. For this definition, the number of molecules in
each shell will change with pressure and temperature~\cite{Schwegle00,scio90}.  
But the orientational measures that most concern us are the tetrahedral arrangement of nearest neighbors,
and bond orientational order in next-nearest neighbors of a central molecule. To see how these orders evolve across
a broad range of state points, we must base our comparison on a fixed number of nearest and
next-nearest neighbors. Moreover, the minima in $g(r)$ become not obvious at high $\rho$, and
g(r) becomes almost featureless beyond the first peak at high $\rho$ (see Fig.~\ref{rdf}(b)).
Hence we choose a less ambiguous shell definition by denoting the nearest four and next-nearest twelve
neighbors of a central water molecule as the first and second shells respectively.

We first study the average effect of density on different shells by dividing $g(r)$ into three regions.
We compute the average number of neighbors of a central molecule at a distance $r$ as
$N(r) \equiv 4 \pi n \int_0^r r'^2 g(r') dr'$, where $n$ is the number density.
We define $r_1$ and $r_2$ such that $N(r_1)=4$ and $N(r_2)=16$.
Therefore, we can define three regions: $0<r \leq r_1$ (first shell),
$r_1<r\leq r_2$ (second shell), and $r>r_2$,
where $r_1$ and $r_2$ depend on $T$ and $\rho$.
Figure~\ref{rdf}(a) shows $N(r)$ at $T=280$~K and $\rho=1.00$~g/cm$^3$ ($n=33.4$/nm$^3$),
where $r_1=0.32$~nm and $r_2=0.48$~nm.

Fig.~\ref{rdf}(b) shows the O-O PCF of TIP5P water at $T=280$~K and
a range of density covering the anomalous regions of water of Fig.~\ref{tmddmqt}. 
Figure~\ref{rdf}(c) shows the change upon compression,
$\Delta g(r) \equiv g(r)|_{\rho} - g(r)|_{\rho_0}$, where $\rho_0=0.88$~g/cm$^3$.
Figure~\ref{rdf}(d) shows the corresponding change, $\Delta N(r) \equiv N(r)|_\rho - N(r)|_{\rho_0}$.
Fig.~\ref{rdf}(b) shows that as $\rho$ increases, the first peak of $g(r)$ decreases,
so $\Delta g(r) <0$ at $r=0.28$~nm in Fig.~\ref{rdf}(c).
This effect of $\rho$ on $g(r)$ is a result of having
a fixed number of neighbors at $r \approx 0.28$~nm, normalized by $n$ in the   
definition of $g(r)$. The change of
the number of neighbors corresponding to the first peak of $g(r)$ is barely
distinguishable (see Fig.~\ref{rdf}(d)), i.e. $\Delta N(r) \approx 0-0.2$ for
$r \approx 0.28$~nm. This implies that the distance defined by the first peak
of $g(r)$ is practically impenetrable and thus, resembles a hard core.
The main changes in $g(r)$ (Fig.~\ref{rdf}(b)) and $\Delta g(r)$ (Fig.~\ref{rdf}(c)) occur in the second
shell. As the density increases, hydrogen bond bending allows water molecules in
the second shell to shift toward the first shell, filling the interstitial space~\cite{soper0002}. 
The changes of $g(r)$ with pressure for $r>r_2$ are minimal.

Figure~\ref{rdf}(d) shows in double logarithmic scale the relationship between $\Delta N(r)$ and $r$.
The slope of curve, $\psi$, characterizes the power law dependence $\Delta N(r) \propto r^\psi$.
There are three main regimes in the behavior of $\Delta N(r)$
as shown by the different slopes $\psi>3$, $\psi<3$ and $\psi=3$. 
The $\psi=3$ at $r > r_2$ is mainly caused by the density change, since $g(r)\approx1$
($\Delta g(r) \approx 0$) for $r > r_2$, so
$\Delta N(r)$ behaves approximately as $\Delta N(r)\propto\Delta\rho~r^3$ with
$\Delta\rho \equiv \rho-\rho_0$. Both the $\psi>3$ and $\psi<3$ regimes are located
within the second shell. The increase of $\Delta N(r)$ for $r < r_2$ is not only due
to density increase, but also due to the shift of water molecules from the second shell
around $0.45$~nm toward the first shell around $0.28$~nm.
Thus, the regime where $\psi>3$, for roughly $r<0.4$~nm, is due to an
increase of $g(r)$ ($\Delta g(r) > 0$), and $\psi<3$ for roughly $0.4$~nm $<r<r_2$
is due to the decrease of $g(r)$ ($\Delta g(r) < 0$). 
Note that $\Delta N(r)=1$ for $r\sim0.33$~nm, corresponding to the fifth neighbor~\cite{scio90}, which is very close 
to the border of the first shell, where $\Delta g(r)$ has its maximum value.
This fifth neighbor in the vicinity of the first shell of water can
produce a defect in the tetrahedral network of water at high density. This defect leads to
hydrogen bond bifurcation and offers paths with low energy barriers between different network
configurations of water. It is also related to diffusion anomaly 
by lowering energy barriers for translational and rotational motions of water molecules~\cite{scio90}.

The translational order parameter $t$ is 
defined in refs.~\cite{truskett00,jeffrey01,shell02,zyan}
\begin{equation}
t\equiv\int_{0}^{s_c}|g(s)-1|ds,
\label{topeq}
\end{equation}
where the dimensionless variable $s \equiv r n^{1/3}$ is the radial distance $r$ scaled
by the mean intermolecular distance $n^{-1/3}$, and $s_c$ usually corresponds
to half of the simulation box size, which is large enough to have
$g(s_c)\approx1$.
We can decompose the translational order parameter $t$ into $t_1, t_2,$ and $t_3$ for each
shell of water by integrating
$|g(s)-1|$ over the three different regions $0<s\leq s_1, s_1<s\leq s_2$
and $s>s_2$, where $s_1=r_1 n^{1/3}$ and
$s_2=r_2 n^{1/3}$. Obviously
\begin{equation}
t=t_1+t_2+t_3.
\end{equation}
The orientational order $q_i$ is used to quantify the tetrahedrality of the first shell,
defined as~\cite{jeffrey01}
\begin{equation}
q_i\equiv1-\frac{3}{8}\sum_{j=1}^{3}\sum_{k=j+1}^{4}\left[\cos\theta_{jik}+\frac{1}{3}\right]^2
\label{qtetra}
\end{equation}
$\theta_{jik}$ is the angle formed between neighbors $j$ and $k$ and
the central molecule $i$.
The average value $q \equiv \frac{1}{N}\sum_{i=1}^{N}q_i$ quantifies the orientational order
of the system based on the molecules in the first shell. 
For perfect tetrahedral order, $q=1$; for an uncorrelated (ideal gas) system, $q=0$.

Because the second shell of the hexagonal ice crystal forms an hcp lattice,
the orientational order parameter for the second shell of water can be characterized by
$Q_{6i}$, which quantifies the extent to which a molecule $i$ and
\emph{twelve} of its neighbors adopt the local fcc, bcc, or hcp structures. 
This orientational order parameter~\cite{steinhardt83} is often used for simple liquids
~\cite{torquato00,truskett00,zyan} because they have
fcc or bcc crystal structures. In order to compute $Q_{6i}$, we first
define twelve bonds connecting each water molecule $i$ with its
twelve next-nearest neighbors in the second shell, and compute for each bond 
its azimuthal and polar angles $(\theta,\varphi)$.
Next we compute $\overline{Y}_{{\ell}m}(\theta,\varphi)$, the average of the
spherical harmonic function over the 12 bonds of the molecule $i$. 
Finally we compute
\begin{equation}
Q_{{\ell}i}\equiv\left[\frac{4\pi}{2\ell+1}\sum_{m=-\ell}^{m=\ell}\vert\overline{Y}_{\ell
m}\vert^2 \right]^\frac{1}{2}.
\label{q6}
\end{equation}
For $\ell=6$, the average value $Q_6 \equiv \frac{1}{N}\sum_{i=1}^{N}Q_{6i}$ quantifies the orientational order
of the system based on the molecules in the second shell.
$Q_{6}$ is large~\cite{steinhardt83} for most crystals such as fcc (0.574),
bcc (0.511), hcp (0.485). For uncorrelated systems, $Q_{6}=1/\sqrt{12}=0.289$.

\begin{figure}
\includegraphics[width=12cm]{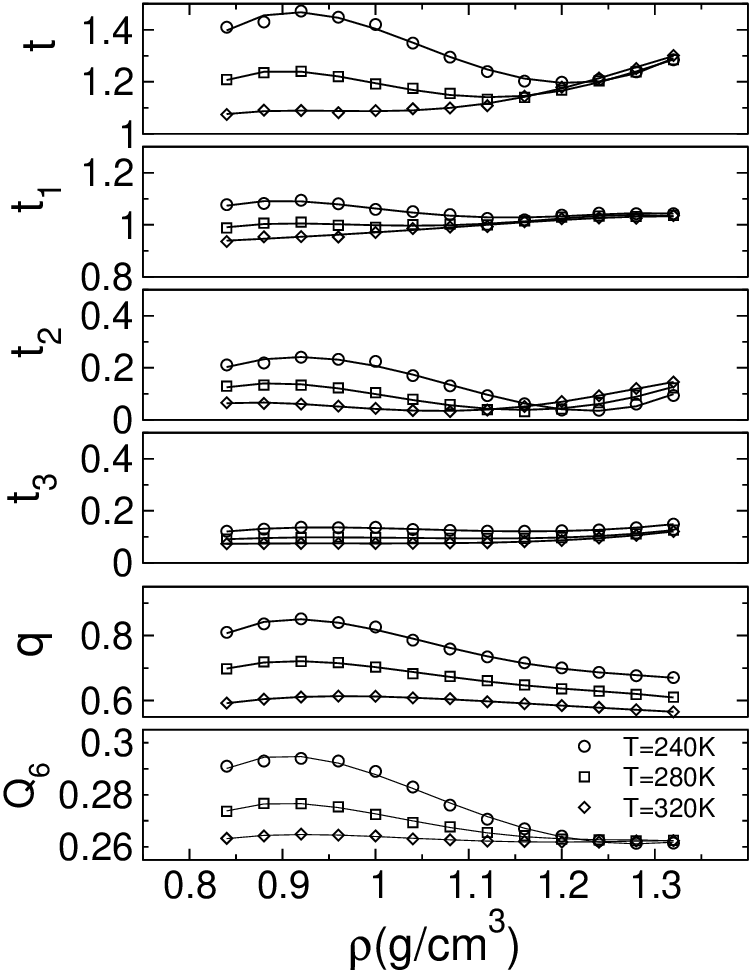}
\caption{
Translational order parameters $t$ (total), $t_1$ (first shell), $t_2$ (second shell), $t_3$
and orientational order parameters $q$ (first shell), $Q_6$ (second shell) of TIP5P water as 
function of density at different $T$. 
The anomalous decrease of orientational order upon compression occurs in both shells ($q, Q_6$),
but the anomalous decrease of translational order upon compression mainly occurs in the second shell ($t_2$).
}
\label{tq-d}
\end{figure}

\begin{figure}
\includegraphics[width=12cm]{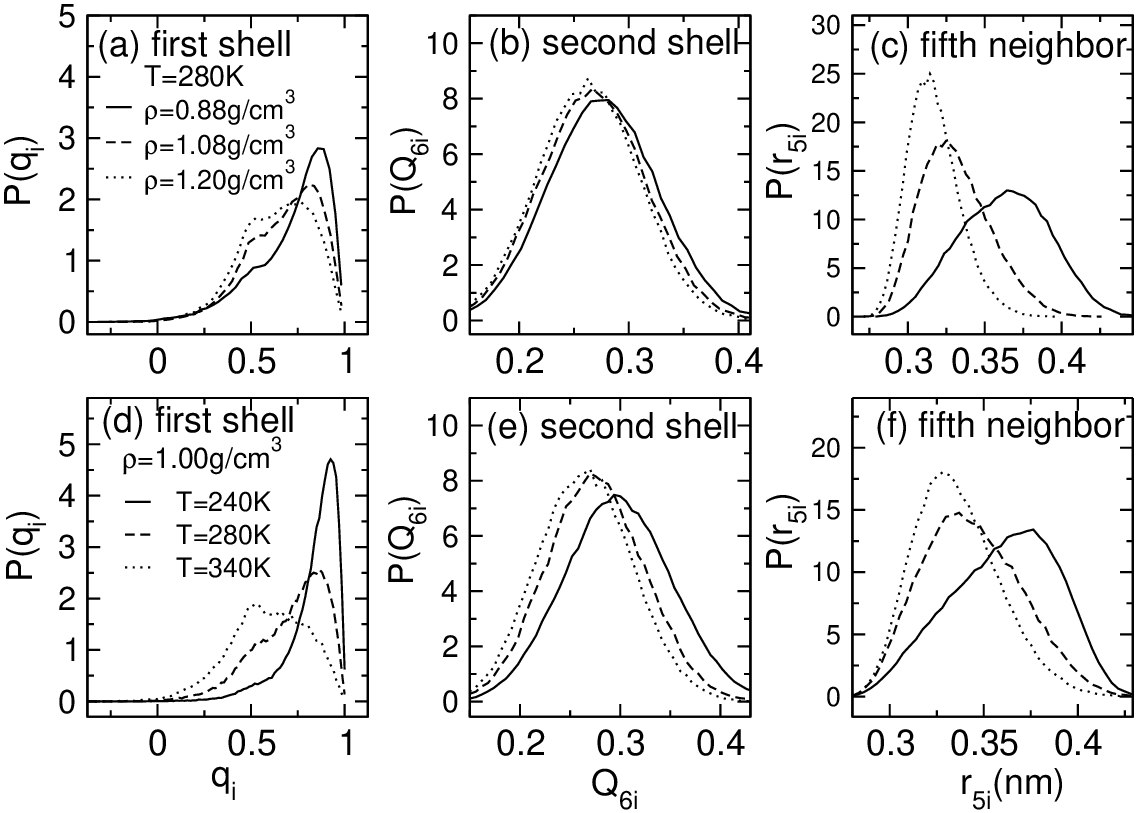}
\caption{
Histograms of (a) the local orientational order $q_i$ in the first shell,
(b) $Q_{6i}$ in the second shell, and (c) distance $r_{5i}$ between
a central water molecule $i$ and its fifth neighbor of TIP5P water.
(a), (b), and (c) show the changes for three different $\rho$ at fixed $T=280K$.
(d), (e), and (f) show the changes for three different $T$ at fixed $\rho=1.00$~g/cm$^3$.
Upon compression or heating over anomalous regions of phase diagram, the fifth neighbor (and other interstitial water
molecules in the second neighbor shell) shift towards first shell (see also Fig.~\ref{rdf} and ref.~\cite{ssitta03}),
corresponding to anomalous changes of structural order in the first and second shells as quantified
by Fig.~\ref{tq-d} and  Fig.~\ref{toprop}.
}
\label{ropqQhis}
\end{figure}

Figure~\ref{tq-d} shows the density dependence of all six order parameters at
three temperatures covering the anomalous
region of TIP5P water (see Fig.~\ref{tmddmqt}). Although $t_1$ is much larger than $t_2$ and $t_3$,
it is apparent that $t_2$ makes the most important contribution to the anomaly
of $t$ (decrease of $t$ with increasing density), compared to $t_1$ and $t_3$.
$t_1$ also makes a small contribution to the $t$ anomaly at low $T=240$~K
due to a small decrease in the first peak of $g(r)$ upon compression.
The anomalous behavior of $t$ becomes weak at $T=280$~K and disappears at $T=320$~K.
The orientational order parameters $q$ and $Q_6$ both show similar anomalous behavior.
The distribution of individual $q_i$ shifts from high $q$ (ice-like) at low $\rho$ and $T$
to low $q$ (less tetrahedral) at high $\rho$ and $T$ as shown in Fig.~\ref{ropqQhis}(a) and (d),
due to increased hydrogen bond bifurcation~\cite{scio90} as interstitial molecules move closely
to the first shell(Fig.~\ref{ropqQhis}(c) and (f)).
$Q_{6i}$ always has approximately normal distribution as shown in Fig.~\ref{ropqQhis}(b) and (e) because there is
no direct bonding between center water molecule and second shell
of water. 

\begin{figure}
\includegraphics[width=12cm]{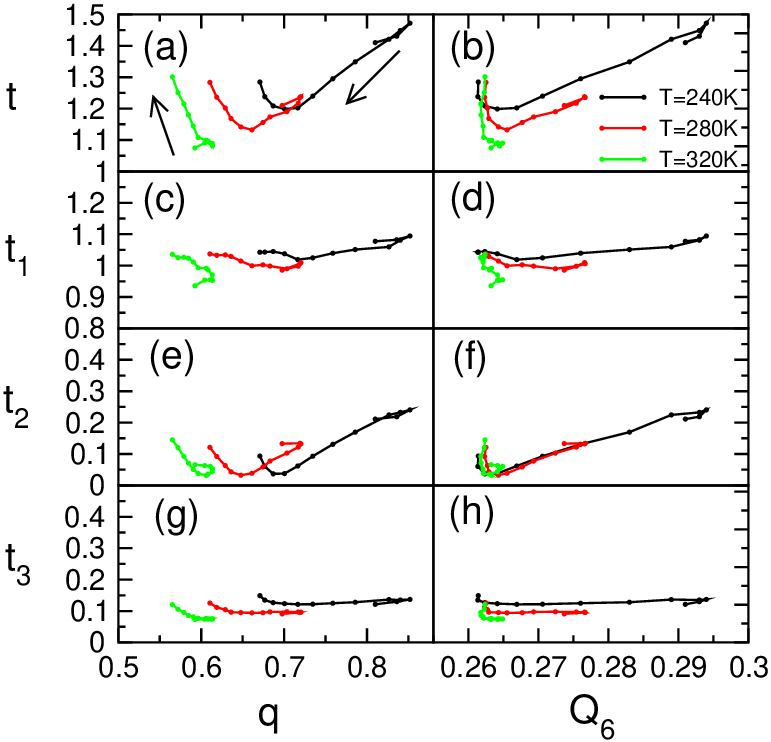}
\caption{
Order maps for TIP5P water (color online).
The arrows indicate the direction of increasing density from $0.84$~g/cm$^3$ to $1.32$~g/cm$^3$.
Only for the second shell order map, $t_2$-$Q_6$ in (f), the isotherms collapse on a line and
the decrease of translational and orientational orders is correlated.
}
\label{toprop}
\end{figure}

A useful way of investigating structural order in fluids is
to map state points onto the $t$-$q$ plane, a representation
called the order map~\cite{torquato00,jeffrey01,zyan}.
The order map for TIP5P water (i.e., using $t$ and $q$) is
shown in Fig.~\ref{toprop}(a). This order map is similar to the
one found in ref.~\cite{jeffrey01} using the SPC/E model. Its main
characteristic is the correlation of the two order parameters
in the anomalous regions where both $q$ and $t$ decrease with density,
as shown by the isotherms collapsing onto a line.
Fig.~\ref{toprop}(b)-(h) shows the
different order maps obtained by considering the order parameters in different shells.
The only one that shows the states in the thermodynamically, dynamically and structurally 
anomalous regions collapsing onto a line, is the panel (f) (i.e. the $t_2$-$Q_6$ order map of the second shell),
indicating that the changes in the second shell are related to anomalies of water.

The first shell order map $t_1$-$q$ in (c) is not correlated
because $t_1$ has only small changes with
increasing density due to the impenetrable hard core at $0.28$~nm, while $q$
changes significantly with density. In the second shell, $t_2$
and $Q_6$ both change significantly and simultaneously with density
so that they are well correlated. Our work quantitatively shows that
the second shell is related to anomalies of water by
its gradual shift towards first shell upon compression.
In addition to water, other tetrahedral liquids such as silica, silicon,
carbon and phosphorous~\cite{angellPCCP} may also exhibit similar
behavior, and a detailed, shell-based study of their order parameters
may prove useful.


\clearpage

\end{document}